\documentstyle[prl,preprint,aps,epsf]{revtex}
\newcommand{\beq}{\begin{equation}}
\newcommand{\eeq}{\end{equation}}
\newcommand{\beqs}{\begin{eqnarray}}
\newcommand{\eeqs}{\end{eqnarray}}
\newcommand{\nn}{\nonumber}

\draft
\begin{document}
\preprint{GTP-97-06}
\title
{Equivalence of renormalization with self-adjoint extension in 
Green's function formalism}
\author
{D.K.Park$^a$ and Sahng-Kyoon Yoo$^b$}
\address
{$^a$ Department of Physics, Kyungnam University, Masan 631-701, Korea \\
$^b$ Department of Physics, Seonam University, Namwon, Chunbuk 590-170, Korea}
\date{\today}
\maketitle
\begin{abstract}
\indent Energy-dependent Green's functions for the two and three dimensional
$\delta$-function plus harmonic oscillator potential systems are derived 
by incorporating the renormalization and the self-adjoint extension into
the Green's function formalism, respectively. It is shown that both methods
yield an identical Green's function if a certain relation between the 
self-adjoint extension parameter and the renormalized coupling constant is 
imposed.  
\end{abstract}


%
%
%
\newpage
Recent study on spin-1/2 Aharonov-Bohm problem[1] and its closely related 
subjects in planar physics[2] produces much attention on a point interaction 
problem in quantum mechanics. In fact, point interaction is solvable from a
mathematical standpoint. For a long time it has been well-known that the 
mathematical method for solving the quantum mechanical 
point interaction problem is the 
self-adjoint extension technique[3], whose essential point is to make the 
Hamiltonian to be self-adjoint operator by extending the Hamiltonian's domain
of definition. In this approach point interaction can be transformed into the
appropriate boundary condition which contains self-adjoint extension 
parameters. 

Although the self-adjoint extension method yields a reasonable solution
on the ground of mathematics, it is evident that more physically 
plausible approach to the point interaction problem is to solve the 
Schr\"{o}dinger equation directly. Two dimensional Schr\"{o}dinger equation for a 
$\delta$-function potential is solved[4], and it has
been found that $\delta$-function potential generates the ultraviolet 
divergence as usual quantum field theories and renormalization procedure
breaks the classical scale symmetry at quantum level. With an aid of the 
scale anomaly 
the two dimensional $\delta$-function potential system 
exhibits the dimensional transmutation phenomenon[5].

Although the ultraviolet divergence is generated by the singular 
$\delta$-function potential and hence, is treated in a similar manner 
to the usual quantum field theories, this simple system has an advantage
that the renormalized exact solution is obtainable. Since our understanding
on the non-perturbative treatment of the ultraviolet divergence is 
incomplete, nonrelativistic $\delta$-function potential system can play 
an important role as a toy model for the study of the non-perturbative 
renormalization scheme. In this context, renormalization procedure and 
renormalization group in the nonrelativistic singular potential systems 
are studied
by many authors[6], recently.

Several years ago comparision of the renormalization with the self-adjoint
extension is examined by R. Jackiw[7] and it has been shown that 
mathematically-based self-adjoint extension and physically-based 
renormalization yield an identical physics in the two and three 
dimensional $\delta$-function potential cases if a certain relation between
the self-adjoint extension parameter and renormalized coupling constant
is imposed. Recently, various point interaction systems 
have been examined in the framework of the 
Green's function formalism[8, 9]. In Ref.[8, 9] present authors have shown
how to incorporate the self-adjoint extension method within the Green's 
function formalism without invoking the perturbation expansion. In the 
present paper we will show that Green's functions for the $\delta$-function
potential systems are also 
obtainable by introducing the appropriate renormalization procedure.
The equivalence of the renormalization with the self-adjoint extension
shown by R. Jackiw in Ref.[7] will be shown again at Green's function 
level by introducing the harmonic oscillator plus $\delta$-function potential
at $ d = 2$ and $d=3$.

Let us start with the two dimensional Hamiltonian
\beq
H = H_0 + v \delta( \vec{r} )
\eeq
where
\beq
H_0 = \frac{\vec{p}^2}{2} + \frac{\omega^2}{2} \vec{r}^2.
\eeq
The harmonic oscillator potential is introduced because of its frequent use
as a calculational tool for the discretization of energy spectrum[10].
By using an expansion
\beq
e^{z \cos \Delta \phi} = \sum_{m = -\infty}^{\infty} I_m(z) e^{i m \Delta 
\phi}
\eeq
where $I_{\nu}(z)$ is usual modified Bessel function, it is straightforward
to derive the energy-dependent Green's function $\hat{G}_0$ for $H_0$:
\beq
\hat{G}_0^{(d = 2)}[\vec{r}_b, \vec{r}_a; E] 
= \sum_{m = - \infty}^{\infty}
\hat{G}_{0, m}^{(d = 2)}[r_b, r_a; E] e^{i m \Delta \theta_{ab}}
\eeq
where
\beqs
\hat{G}_{0, m}^{(d = 2)}[r_b, r_a; E]  
&=& \frac{1}{2 \pi \omega r_a r_b}
    \frac{\Gamma \left( \frac{1 + m + E / \omega}{2} \right) }
         {\Gamma (1 + m)}  \\  \nn
&\times&    W_{- \frac{E}{2 \omega}, \frac{m}{2}} (\omega Max(r_a^2, r_b^2))
    M_{- \frac{E}{2 \omega}, \frac{m}{2}} (\omega Min(r_a^2, r_b^2)).
\eeqs
Here, $W_{\kappa, \mu}(z)$ and $M_{\kappa, \mu}(z)$ are usual Whittaker
functions. Since $\delta$-function potential is zero range, so it gives 
some modification only on s-wave, the difference of energy-dependent
Green's functions for $H$ and for $H_0$ is independent of the 
polar angle $\theta$. Hence, one can define the difference of the 
energy-dependent Green's functions as follows:
\beqs
\hat{A}^{(d = 2)} [r_b, r_a; E]  
&\equiv& \hat{G}^{(d = 2)}[\vec{r}_b, \vec{r}_a; E] - 
         \hat{G}_0^{(d = 2)}[\vec{r}_b, \vec{r}_a; E]  \\  \nn
&=& \hat{G}_{m = 0}^{(d = 2)}[r_b, r_a; E] - 
    \hat{G}_{0, m=0}^{(d = 2)}[r_b, r_a; E]
\eeqs
where $\hat{G}^{(d = 2)}[\vec{r}_b, \vec{r}_a; E]$ is energy-dependent 
Green's function for $H$ and $\hat{G}_{m = 0}^{(d = 2)}[r_b, r_a; E]$ is 
radial part of s-wave in $\hat{G}^{(d = 2)}[\vec{r}_b, \vec{r}_a; E]$.
By following the method used in Ref.[8] one can express
$\hat{A}^{(d=2)}[r_b, r_a; E]$ in terms of 
$\hat{G}_{0, m=0}^{(d = 2)}[r_b, r_a; E]$ :
\beq
\hat{A}^{(d=2)}[r_b, r_a; E]  
=- \frac{\hat{G}_{0, m=0}^{(d = 2)}[r_b, \epsilon_1; E]
          \hat{G}_{0, m=0}^{(d = 2)}[\epsilon_2, r_a; E]}
         {\frac{1}{v} + \lim_{\epsilon_2 \rightarrow \epsilon_1^+}
          \hat{G}_{0, m=0}^{(d = 2)}[\epsilon_2, \epsilon_1; E]}.
\eeq
In Eq.(7) the limit $\epsilon_1 \rightarrow 0$ should be taken 
after calculation.
By combining the Eqs.(5) and (7) and using the relations of the Whittaker
functions with the confluent hypergeometric functions[11]
\beqs
M_{\kappa, \mu}(z)&=& e^{-\frac{z}{2}} z^{\frac{1}{2} + \mu}
                      M(\frac{1}{2} + \mu - \kappa, 1 + 2 \mu; z) \\ \nn
W_{\kappa, \mu}(z)&=& e^{-\frac{z}{2}} z^{\frac{1}{2} + \mu}
                      U(\frac{1}{2} + \mu - \kappa, 1 + 2 \mu; z),
\eeqs
more convenient form of $\hat{A}^{(d=2)}[r_b, r_a; E]$ is 
\beq
\hat{A}^{(d=2)}[r_b, r_a; E] = f_0(\epsilon_1, \epsilon_2; E)
                               g_0(r_a) g_0(r_b)
\eeq
where
\beqs
f_0(\epsilon_1, \epsilon_2; E)&=&
- \frac{1}
       {\frac{1}{v} + \lim_{\epsilon_2 \rightarrow \epsilon_1^+}
          \hat{G}_{0, m=0}^{(d = 2)}[\epsilon_2, \epsilon_1; E]}
  \frac{1}{(2 \pi \omega)^2 \epsilon_1 \epsilon_2}  \\  \nn
&\times& \Gamma^2 \left( \frac{1}{2} + \frac{E}{2 \omega} \right)
         M_{- \frac{E}{2 \omega},0}(\omega \epsilon_1^2)
         M_{- \frac{E}{2 \omega},0}(\omega \epsilon_2^2)
\eeqs
and
\beq
g_0(r) = \sqrt{\omega} e^{- \frac{\omega r^2}{2}}
         U(\frac{1}{2} + \frac{E}{2 \omega}, 1; \omega r^2).
\eeq
Here, $f_0(\epsilon_1, \epsilon_2; E)$ has an ultraviolet divergence 
because of the denominator term
$\frac{1}{v} + \lim_{\epsilon_2 \rightarrow \epsilon_1^+}
 \hat{G}_{0, m=0}^{(d = 2)}[\epsilon_2, \epsilon_1; E]$. 

In the following we will show how to make the theory finite by introducing
the renormalized coupling constant, which is assumed to be a finite quantity.
Using the asymptotic formula
\beqs
\lim_{z \rightarrow 0} M(a, 1, z) &=& 1,   \\   \nn
\lim_{z \rightarrow 0} U(a, 1, z) &=& - \frac{1}{\Gamma(a)}
                                      [\ln z + \psi(a)]
\eeqs
where $\psi(z)$ are digamma function, 
the denominator in Eq.(7)
becomes
\beq
\frac{1}{v} + \lim_{\epsilon_2 \rightarrow \epsilon_1^+}
\hat{G}_{0, m=0}^{(d = 2)}[\epsilon_2, \epsilon_1; E] 
= \frac{1}{v} - \frac{1}{2 \pi}
\left[ \ln \frac{\omega}{\mu} + \ln \mu \epsilon_2^2 + 
       \psi \left( \frac{1}{2} + \frac{E}{2 \omega} \right)  \right]
\eeq
where arbitrary mass-dimensional parameter $\mu$ is introduced.

In order to make the theory finite we introduce the renormalized 
coupling constant $g$
\beq
\frac{1}{g} \equiv \frac{1}{v} - \frac{1}{2 \pi} \ln \mu \epsilon_2^2,
\eeq
which leads Eq.(13) to
\beq
\frac{1}{v} + \lim_{\epsilon_2 \rightarrow \epsilon_1^+}
\hat{G}_{0, m=0}^{(d = 2)}[\epsilon_2, \epsilon_1; E] 
= \frac{1}{g} - \frac{1}{2 \pi}
    \left[ \ln \frac{\omega}{\mu} + 
           \psi \left( \frac{1}{2} + \frac{E}{2 \omega} \right) \right].
\eeq
Inserting Eq.(15) into Eq.(10) and using the relations (8) again one can easily
obtain the finite $f_0(\epsilon_1, \epsilon_2; E)$
\beq
f_0(\epsilon_1, \epsilon_2; E)
= \frac{\Gamma^2 \left( \frac{1}{2} + \frac{E}{2 \omega} \right)}
       {4 \pi^2 \omega}
  \frac{1}
       {\frac{1}{2 \pi} \left[ \ln \frac{\omega}{\mu} +
        \psi \left( \frac{1}{2} + \frac{E}{2 \omega} \right) \right]
        - \frac{1}{g}}.
\eeq
By combining Eqs.(9), (11), and (16) 
$\hat{A}^{(d=2)}[r_b, r_a; E]$ becomes finite as follows:
\beq
\hat{A}^{(d=2)}[r_b, r_a; E] 
= \frac{\Gamma^2 \left( \frac{1}{2} + \frac{E}{2 \omega} \right)}
         {4 \pi^2 \omega r_a r_b}
    \frac{1}
       {\frac{1}{2 \pi} \left[ \ln \frac{\omega}{\mu} +
        \psi \left( \frac{1}{2} + \frac{E}{2 \omega} \right) \right]
        - \frac{1}{g}}
    W_{-\frac{E}{2 \omega}, 0} (\omega r_a^2)
    W_{-\frac{E}{2 \omega}, 0} (\omega r_b^2).
\eeq

Now let us apply the self-adjoint extension method to derive the finite
$\hat{A}^{(d=2)}[r_b, r_a; E]$. Following Ref.[8] it is straightforward to 
show that the boundary condition
\beq
\lim_{r \rightarrow 0} \frac{g_0(r)}{\ln r}
= \frac{\lambda}{\pi} \lim_{r \rightarrow 0}
 \left[ g_0(r) - \left( \lim_{r^{\prime} \rightarrow 0}
                        \frac{g_0(r^{\prime}}{\ln r^{\prime}}  \right)
        \ln r  \right],
\eeq
where $\lambda$ is real self-adjoint extension parameter and $g_0(r)$ is 
defined in Eq.(11), is derived by incorporating the self-adjoint extension
method into the Green's function formalism properly. As a result of 
the boundary
condition (18) the condition for the bound-state spectrum
\beq
\ln \omega + \psi \left( \frac{1}{2} + \frac{E}{2 \omega} \right)
= \frac{2 \pi}{\lambda}
\eeq
is easily obtained. Since bound-state energy must be a pole of the 
energy-dependent Green's function, we get another condition for the 
bound-state spectrum
\beq
\frac{1}{v} - \frac{1}{2 \pi}
    \left[ \ln \omega \epsilon_2^2 + 
           \psi \left( \frac{1}{2} + \frac{E}{2 \omega} \right) \right] = 0.
\eeq
In order for the conditions (19) and (20)
to be consistent the relation between the 
self-adjoint extension parameter $\lambda$ and the bare coupling 
constant $v$
\beq
\frac{1}{v} = \frac{1}{\lambda} + \frac{1}{\pi} \ln \epsilon_2
\eeq
must be required.
The remaining calculation procedure to obtain the 
$\hat{A}^{(d=2)}[r_b, r_a; E]$ is exactly same with the previous 
renormalization method.
Finally, we obtain
\beq
\hat{A}^{(d=2)}[r_b, r_a; E]   
=
\frac{\Gamma^2 \left( \frac{1}{2} + \frac{E}{2 \omega} \right)}
         {4 \pi^2 \omega r_a r_b}
\frac{1}
     {\frac{1}{2 \pi} \left[ \ln \omega + 
                             \psi \left( \frac{1}{2} + \frac{E}{2 \omega} \right)
                                            \right]
        - \frac{1}{\lambda} }
      W_{-\frac{E}{2 \omega}, 0} (\omega r_a^2)
    W_{-\frac{E}{2 \omega}, 0} (\omega r_b^2).
\eeq
By comparing Eq.(14) with Eq.(21) the relation between the self-adjoint
extension parameter $\lambda$ and the renormalized coupling constant $g$
\beq
\frac{1}{\lambda} = \frac{1}{g} + \frac{1}{2 \pi} \ln \mu
\eeq
is simply derived. Hence, the self-adjoint extension
and renormalization methods yield an identical energy-dependent Green's
function if the relation (23) is assumed. The final form of 
$\hat{G}^{(d=2)}[\vec{r}_b, \vec{r}_a; E]$ is 
\beq
\hat{G}^{(d=2)}[\vec{r}_b, \vec{r}_a; E]
= \sum_{m = - \infty}^{\infty} \hat{G}_m^{(d=2)}[r_b, r_a; E]
  e^{i m \theta_{ab}}
\eeq
where
\beqs
\hat{G}_{m \neq 0}^{(d=2)}[r_b, r_a; E]&=& 
\hat{G}_{0, m \neq 0}^{(d=2)}[r_b, r_a; E]  \\  \nn
\hat{G}_{m=0}^{(d=2)}[r_b, r_a; E]&=&
\hat{G}_{0, m=0}^{(d=2)}[r_b, r_a; E] + \hat{A}^{(d=2)}[r_b, r_a; E].
\eeqs

The equivalence of the self-adjoint extension and the renormalization 
is also straightforwardly shown in the three dimensional $\delta$-function
plus harmonic oscillator system by following the exactly same procedure.
In this case the final form of 
$\hat{G}^{(d=3)}[\vec{r}_b, \vec{r}_a; E]$ is 
\beq
\hat{G}^{(d=3)}[\vec{r}_b, \vec{r}_a; E] = 
\sum_{l=0}^{\infty} \sum_{m = -l}^{l}
\hat{G}_l^{(d=3)}[r_b, r_a; E]
Y_{l,m}^{\ast}(\theta_a, \phi_a) Y_{l,m}(\theta_b, \phi_b)
\eeq
where
\beqs
& &\hat{G}_{l=0}^{(d=3)}[r_b, r_a; E]  \\  \nn       
&=& \frac{1}{\omega (r_a r_b)^{3/2}}
    \frac{\Gamma \left( \frac{3}{4} + \frac{E}{2 \omega} \right)}
         {\Gamma \left( \frac{3}{2} \right)}
    W_{-\frac{E}{2 \omega}, \frac{1}{4}} (\omega Max(r_a^2, r_b^2))
    M_{-\frac{E}{2 \omega}, \frac{1}{4}} (\omega Min(r_a^2, r_b^2))  \\  \nn
&+& \frac{1}{\pi \sqrt{\omega} (r_a r_b)^{3/2}}
    \frac{\Gamma \left( \frac{3}{4} + \frac{E}{2 \omega} \right)}
         { \frac{\sqrt{\omega}} 
                {\Gamma \left( \frac{1}{4} + \frac{E}{2 \omega} \right) }
           - \frac{\pi / \lambda}
                  {\Gamma \left( \frac{3}{4} + \frac{E}{2 \omega} \right)} }
    W_{-\frac{E}{2 \omega}, \frac{1}{4}} (\omega r_a^2)
    W_{-\frac{E}{2 \omega}, \frac{1}{4}} (\omega r_b^2),  \\   \nn
& &\hat{G}_{l \neq 0}^{(d=3)}[r_b, r_a; E]  \\  \nn
&=& \frac{1}{\omega (r_a r_b)^{3/2}}
    \frac{ \Gamma\left( \frac{3}{4} + \frac{l}{2} + \frac{E}{2 \omega} \right)}
         {\Gamma \left( l + \frac{3}{2} \right)}
    W_{-\frac{E}{2 \omega}, \frac{l}{2} + \frac{1}{4}} (\omega Max(r_a^2, r_b^2))
    M_{-\frac{E}{2 \omega}, \frac{l}{2} + \frac{1}{4}} (\omega Min(r_a^2, r_b^2)
)
\eeqs
and $Y_{l,m}(\theta, \phi)$ is usual spherical harmonics.
In this case $\lambda$, real self-adjoint extension parameter, is related 
to the renormalized coupling constant $g$ as follows:
\beq
\frac{1}{g} = \frac{4 \pi}{\lambda}
\eeq
where $\frac{1}{g} = \frac{1}{v} + \frac{1}{\epsilon_2}$.

If one takes $\omega \rightarrow 0$ limit in Eq.(24) and Eq.(26),  
the same energy-dependent Green's functions
with those given in Ref.[8] are straightforwardly derived 
by using the various asymptotic formulae.

In summary, the energy-dependent Green's functions for the two and three
dimensional $\delta$-function plus harmonic oscillator systems are 
derived by using the renormalization and self-adjoint extension 
methods, respectively. It is shown that the Green's function derived
by each method coincides with each other 
if the relations between the self-adjoint extension parameter and the 
renormalized coupling constant, Eq.(23) at $d=2$ and Eq.(28) at $d=3$
are imposed.



\normalsize
\newpage
\begin{thebibliography}{99}
\bibitem{1} Ph.de Sousa Gerbert, Phys. Rev. {\bf D40}, 1346 (1989); \\
C.R.Hagen, Int. J. Mod. Phys. {\bf A6}, 3119 (1991).
\bibitem{2} R.Jackiw, Nucl. Phys. (Proc. Suppl.) {\bf A18}, 107 (1990); \\
M.G.Alford, J.March-Russell, and F.Wilczek, Nucl. Phys. {\bf B328}, 140 (1989); \\
Y.H.Chen, F.Wilczek, E.Witten, and B.I.Halperin, Int. J. Mod. Phys. {\bf B3}, 
1001 (1989).   
\bibitem{3} S.Albeverio, F.Gesztesy, R.Hoegh-Krohn and H.Holden, {\it
Solvable Models in Quantum Mechanics}, (Springer-Verlag, Berlin, 1988).
\bibitem{4} C.Thorn, Phys. Rev. {\bf D19}, 639 (1979); \\
K. Huang, {\it Quarks, Leptons, and Gauge Fields}, (World Scientific, 
Singapore, 1982).
\bibitem{5} S.Coleman and E.Weinberg, Phys. Rev. {\bf D7}, 1888 (1973).
\bibitem{6} C.Manuel and R.Tarrach, Phys. Lett. {\bf B328}, 113 (1994); \\
S.K.Adhikari, T.Frederico, and I.D.Goldman, Phys. Rev. Lett. {\bf 74}, 487 
(1995);  \\
S.K.Adhikari and T.Frederico, Phys. Rev. Lett. {\bf 74}, 4572 (1995);  \\
R.J.Henderson and S.G.Rajeev, J. Math. Phys. {\bf 38}, 2171 (1997); 
hep-th/9710061;  \\
S.K.Adhikari and A.Ghosh, J. Phys. {\bf A30}, 6553 (1997);  \\
D.R.Phillips, S.R.Beane, and T.D.Cohen, hep-th/9706070.  
\bibitem{7} R.Jackiw, {\it in M.A.B\'{e}g Memorial Volume}, edited by
A.Ali and P.Hoodbhoy (World Scientific, Singapore, 1991).
\bibitem{8} D.K.Park, J. Math. Phys. {\bf 36}, 5453 (1995).
\bibitem{9} D.K.Park, J. Phys. A: Math. Gen. {\bf 29}, 6407 (1996);  \\
D.K.Park and Shang-Kyoon Yoo, to appear in Ann. Phys. 
\bibitem{10} A.Comtet, Y.Georgelin, and S.Ouvry, J. Phys. {\bf A22},
3917 (1989);  \\ 
T.Blum, C.R.Hagen and S.Ramaswamy, Phys. Rev. Lett. {\bf 64}, 
709 (1990).
\bibitem{11} L.J.Slater, {\it in Handbook of Mathematical Functions with
Formulas, Graphs, and Mathematical Tables}, edited by 
M.Abramowitz and I.A.Stegun (Dover, New York, 1972).
\end{thebibliography}
\end{document}